\documentclass[aps,prl,twocolumn,superscriptaddress]{revtex4}
\usepackage{amsfonts}
\usepackage{amsmath}
\usepackage{amssymb}
\usepackage{graphicx}

\begin{document}
\title{
How do packing defects modify the cooperative motions in supercooled liquids ?
}

\author{Sonia Taamalli}
\affiliation{ Laboratoire de Photonique d'Angers EA 4464, Universit\' e d'Angers, Physics Department,  2 Bd Lavoisier, 49045 Angers, France}
\affiliation{ University of Monastir, Physics Department,  Monastir, Tunisia}

\author{Hafedh Belmabrouk}
\affiliation{ University of Monastir, Physics Department,  Monastir, Tunisia}

\author{Vo Van Hoang}
\affiliation{ University of Ho Chi Minh, Physics Department,  Ho Chi Minh City, Vietnam}

\author{Victor Teboul}
\email{victor.teboul@univ-angers.fr}
\affiliation{ Laboratoire de Photonique d'Angers EA 4464, Universit\' e d'Angers, Physics Department,  2 Bd Lavoisier, 49045 Angers, France}


\begin{abstract}

We use molecular dynamic simulations to investigate the relation between the presence of packing defects in a glass-former and the spontaneous cooperative motions called dynamic heterogeneity. For that purpose we use a simple diatomic glass-former and add a small number of larger or smaller diatomic probes. The diluted probes modify locally the packing, inducing structural defects in the liquid, while we find that the number of defects is small enough not to disturb the average structure. We find that a small packing modification around a few molecules can deeply influence the dynamics of the whole liquid, when supercooled. When we use small probe molecules, the dynamics accelerates and the dynamic heterogeneity decreases. In contrast, for large probes the dynamics slows down and the dynamic heterogeneity increases. The induced heterogeneities and transport coefficient modification increase when the temperature decreases and disappear around the onset temperature of the cage dynamics.

\end{abstract}

\maketitle
\section{ Introduction}

The spontaneous appearance of cooperative motions called dynamic heterogeneity (DH) when liquids are supercooled \cite{dh1,dh2,dh3,dh4,dh5,strings,strings2,strings3,gt1} has been extensively studied during the last decade, as scientists were searching for a cooperative mechanism at the origin of the glass-transition.
Despite large efforts in that direction of research however,  the origin of the dynamic heterogeneity and the effect of these cooperative motions on the liquid dynamics still remain elusive. 
It is widely thought that DHs appear due to the presence of some defects inside the low temperature liquid, that could be structural (a local packing fluctuation) or purely dynamical (excitations).  The resulting excitations (arising directly or from structural defects) then grow due to facilitation\cite{facilitation,chandler,defect} processes, eventually disappearing at large timescales due to thermal diffusion.
In 2004, Asaph Widmer-Cooper and coworkers\cite{widmer} showed that DHs appear preferentially in some regions of the liquid, a result that strongly suggests a connection between the local structure and the DHs.
In 2007, Laura Kaufman's group \cite{lk1,lk2} studied the effect of probes larger than the medium molecules on the dynamics of supercooled liquids. They found that the dynamics slows down or accelerates depending on the probe's roughness, while the DHs  increase in both cases, and concluded that probes can promote dynamic heterogeneities. 
In 2009, we found that dynamical defects created by probes isomerizing inside the medium can strongly promote DHs\cite{us1,us2} while accelerating the dynamics, results that were confirmed by experiments and simulations\cite{c1,c2,c3,c4,c5}.
Following these works, results on colloids\cite{weeks,coll2,coll3,coll4,coll5}, and theoretical findings\cite{vk1,vk2,vk3}, in order to test the hypothesis of a packing fluctuation origin of the DHs, we investigate in this paper the effect of small structural defects on the dynamics of a model supercooled liquid and on the strength of dynamic heterogeneity.
We focus our work on probes of sizes comparable with  the size of the medium's molecules in order to model small cages fluctuations.
We also focus on the possibility for  some probes to destroy the dynamic heterogeneities.
In agreement with the results of Kaufman's group we find that large probes promote the DHs in our liquid, however we also find that small probes can destroy the heterogeneities.

\section{Calculation}

In this work we use molecular dynamics simulations\cite{md1,md2,md3} (MD) to simulate the dynamics of a fragile\cite{fragile1,fragile2} supercooled liquid when small or large probe molecules are diluted inside. This simulation method permits  to gain information on the motion of each molecule of the medium provided that the interatomic potentials are known with enough accuracy, and is thus an invaluable tool to unravel condensed matter physics phenomena\cite{mdcm-1,mdcm-2,mdcm-3,mdcm-4,mdcm0,mdcm1,mdcm2} at the microscopic level. 
We model the molecules\cite{ariane} of the medium as constituted of two atoms ($i=1, 2$) that we rigidly bond fixing the interatomic distance to $d^{m}=2.28 $\AA$ $.
 Each atom of our linear molecule has a mass $m^{m}=50$ g/$N_{a}$ where $N_{a}$ is the Avogadro number.
Atoms of the set of linear molecules constituting our liquid interact with the following Lennard-Jones potentials: 
\begin{equation}
V_{ij}=4\epsilon_{ij}((\sigma_{ij}/r)^{12} -(\sigma_{ij}/r)^{6}) 
\end{equation}

 with the parameters: $\epsilon_{11}^{m}= \epsilon_{12}^{m}=0.5$ kJ/mol, $\epsilon_{22}^{m}= 0.4$ kJ/mol, $\sigma_{11}^{m}= \sigma_{12}^{m}=4.56$\AA\ and $\sigma_{22}^{m}=4.33$\AA$ $.
   Our mean molecule is thus $6.7$\AA$ $ long and $4.5$\AA$ $ wide.
With these parameters the liquid does not crystallize when supercooled even during long simulation runs\cite{ariane}. This model has been described and studied in detail previously\cite{ariane,finitesize} and was found to display the typical behaviors of fragile supercooled liquids.
 As they are modeled with Lennard-Jones atoms, the potentials are quite versatile.
Due to that property, a shift in the parameters $\epsilon$ will shift all the temperatures by the same amount, including the glass-transition temperature and the melting temperature of the material.
We add to that medium $1\%$ smaller or larger molecules (the probes) intended to create small localized perturbations in the medium structure.   We have $6$ probe molecules diluted within $600$ medium molecules inside the simulation box.
This number of molecules lead to a simulation box size of $39.65$\AA, that insures the disappearance of size effects\cite{finitesize,size} in our liquid at the temperature of study.
The probes are similar diatomic molecules defined by the parameters: 
$\epsilon_{ij}^{L}= \epsilon_{ij}^{S}= \epsilon_{ij}^{m}$, $\sigma_{ij}^{L}= \sigma_{ij}^{m}$, $\sigma_{12}^{S}= \sigma_{12}^{m}$, $\sigma_{11}^{S}=2.28$\AA, $\sigma_{22}^{S}=2.17$\AA$ $, $m^{L}=m^{S}=m^{m}$,  $d^{L}=5.72 $\AA\  and $d^{S}=0.47 $\AA.
The density is set constant in our calculations at $\rho=1.615 g/cm^{3}$. When rescaled, or in dimensionless units, that density value is larger than the density of the original model\cite{ariane}, and thus leads to a more viscous medium. 
We use the Gear algorithm with the quaternion method\cite{md1} to solve the equations of motions with a time step $\Delta t=2$ $10^{-15} s$ above $T=120K$ and $\Delta t=4 $ $10^{-15} s$ below that temperature. 
The temperature is controlled using a Berendsen thermostat\cite{berendsen}. To prevent aging processes we equilibrate the supercooled liquid during $100$ ns, a time much larger than the $\alpha$ relaxation time at the lower temperature studied, before recording the simulation results. We use periodic boundary conditions.

\section{Results and discussion}


We  expect slower motions around the large inclusions and more rapid motions around the small inclusions.
This modification of the molecular motions around the defects will result in a local heterogeneity of the mobilities, that may induce or destroy dynamic heterogeneities if facilitation mechanisms are present in the system.
We thus expect a modification of the liquid's dynamic heterogeneity around the inclusions.

\includegraphics[scale=0.3]{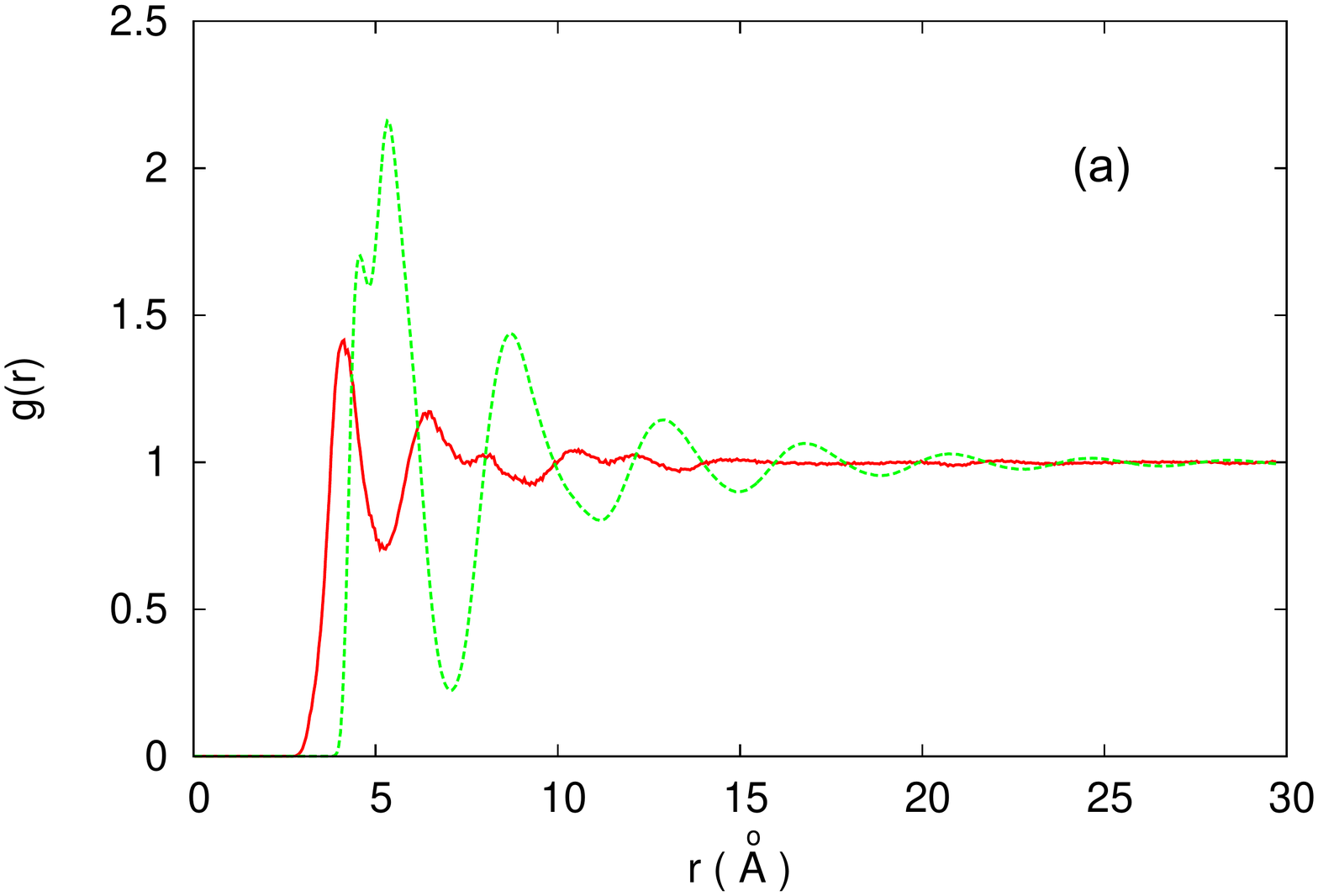}

\includegraphics[scale=0.3]{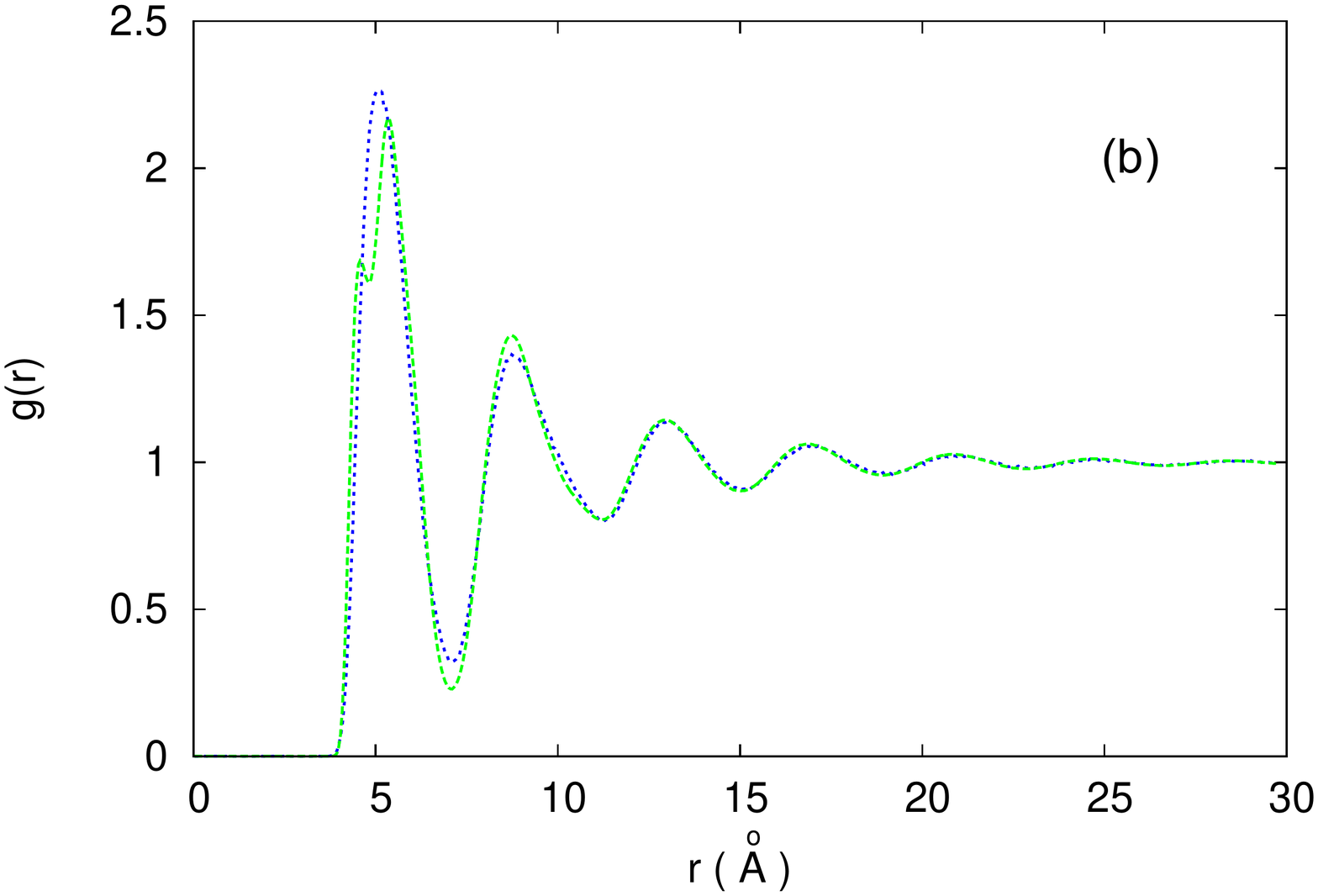}

{\em \footnotesize  FIG.1. (color online)  Radial distribution function $g(r)$ between the centers of masses (CoM) of the medium molecules (dashed blue line) and between the average molecules and the defects (continuous red line).
(a) Large defects; (b)  Small  defects. The $g(r)$ between the average molecules doesn't change in our simulations when defects are included, due to the small proportion of defects. 
The temperature is T=120K.\\}

Due to the small percentage of probes, we do not find any modification of the mean structure of the liquid when we replace a few ($1\%$) molecules of the medium by larger or smaller probe molecules.
The local structure is however modified around the probe  as shown in Figure 1a and 1b.
Figure 1 suggests that the cage is smaller around the large probes and  slightly larger around the small probe, as the first peak of the radial distribution function is shifted to a smaller distance for large probes than for medium molecules.
The Figure also shows that the local structure around the probe is only slightly modified when the small probes are used.

\includegraphics[scale=0.3]{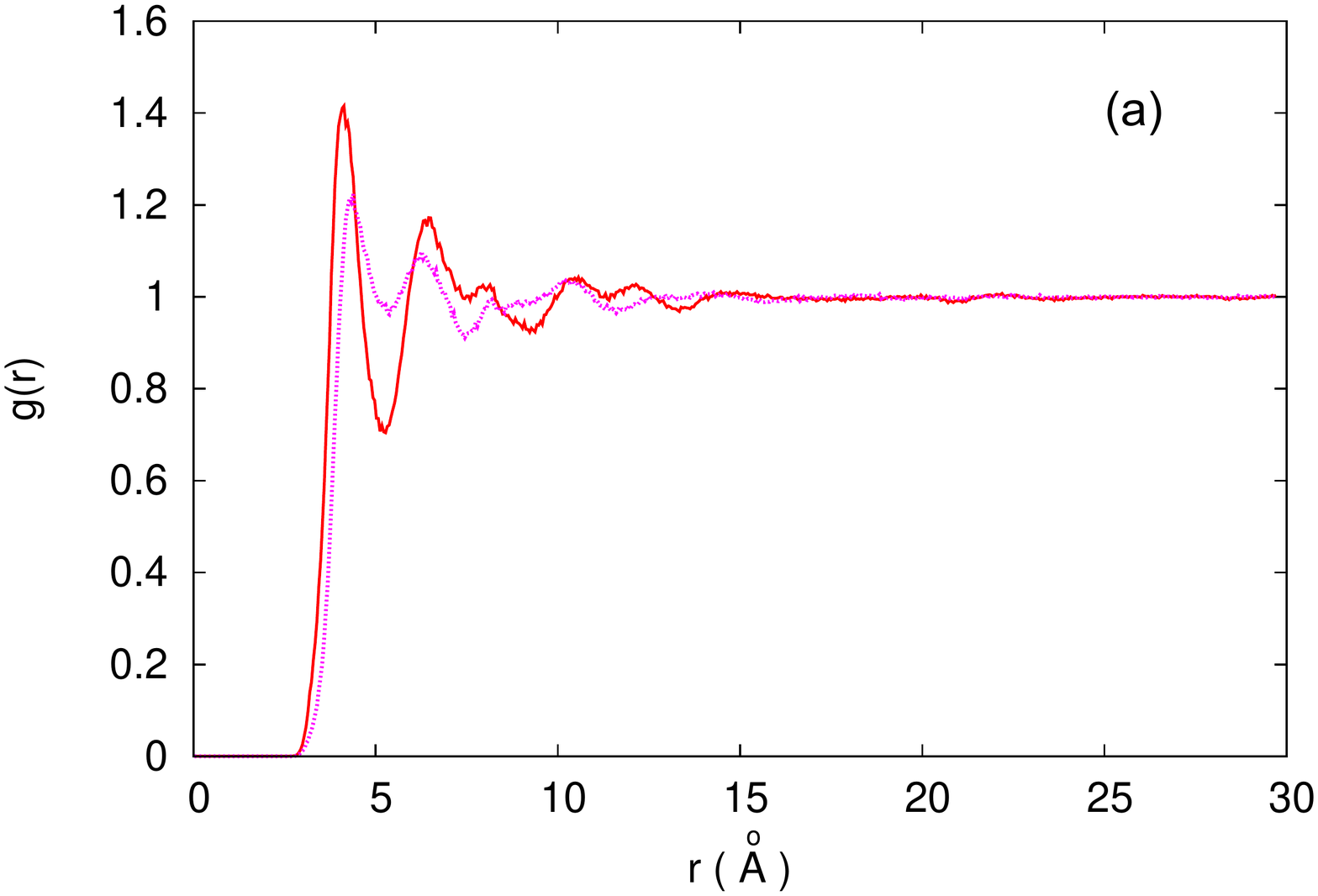}

\includegraphics[scale=0.3]{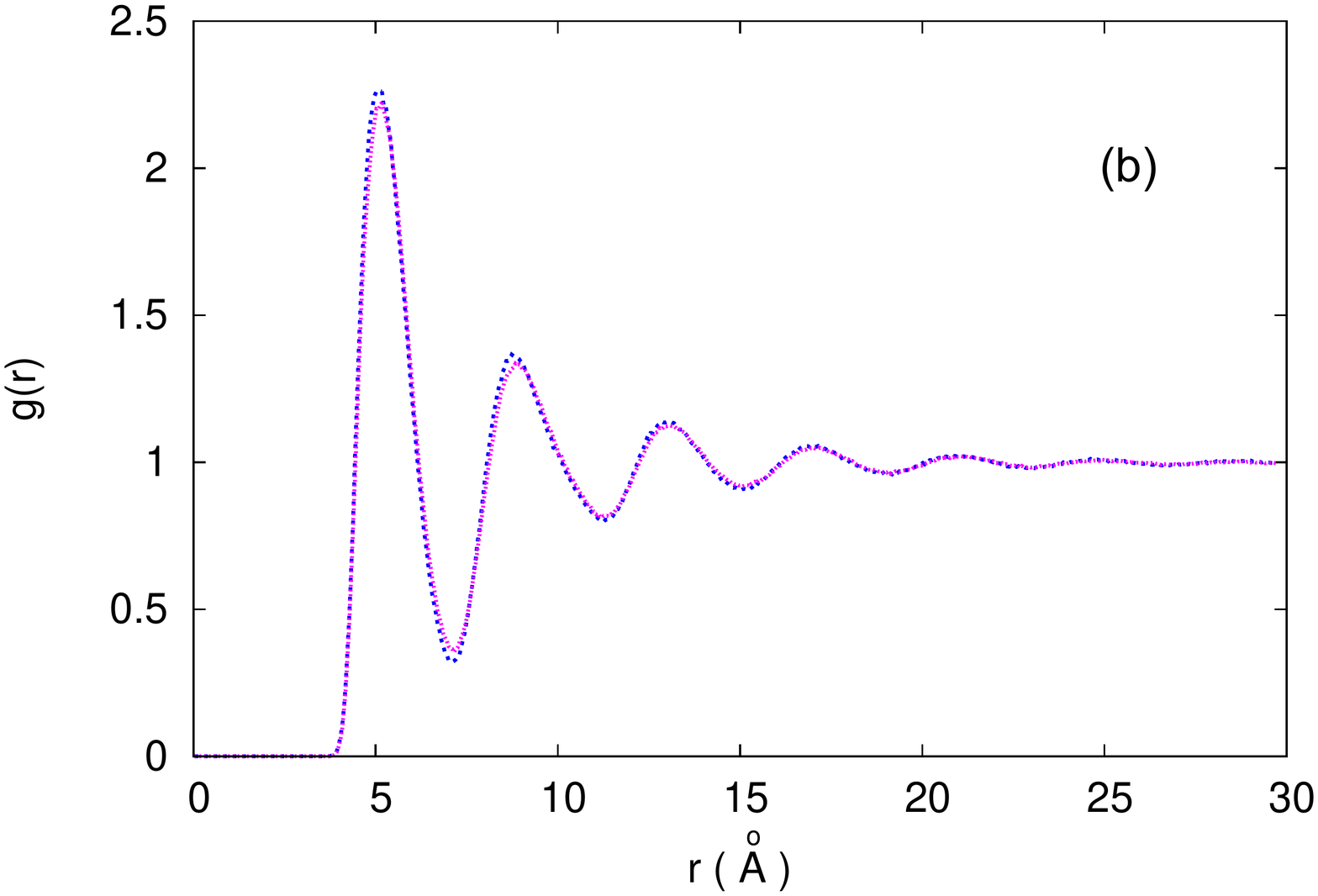}

{\em \footnotesize  FIG.2. (color online)  Radial distribution function $g(r)$ between the centers of masses (CoM) of the medium molecules and of the defects (continuous red line) at a temperature T=120K (continuous red line) and T=165K (dashed green line).
(a) Large defects; (b)  Small  defects.\\}

Figure 2 shows that the temperature modifies  the structure around the large probes,  increasing the probability of unfavored geometrical configurations. 
By contrast, for small probes the structure remains the same around the probes at low and high temperature.
However Figures 3 and 4 show that both probes modify the dynamics. This modification of the dynamics begins around $T\approx200K$ and increases when the temperature drops.

\includegraphics[scale=0.3]{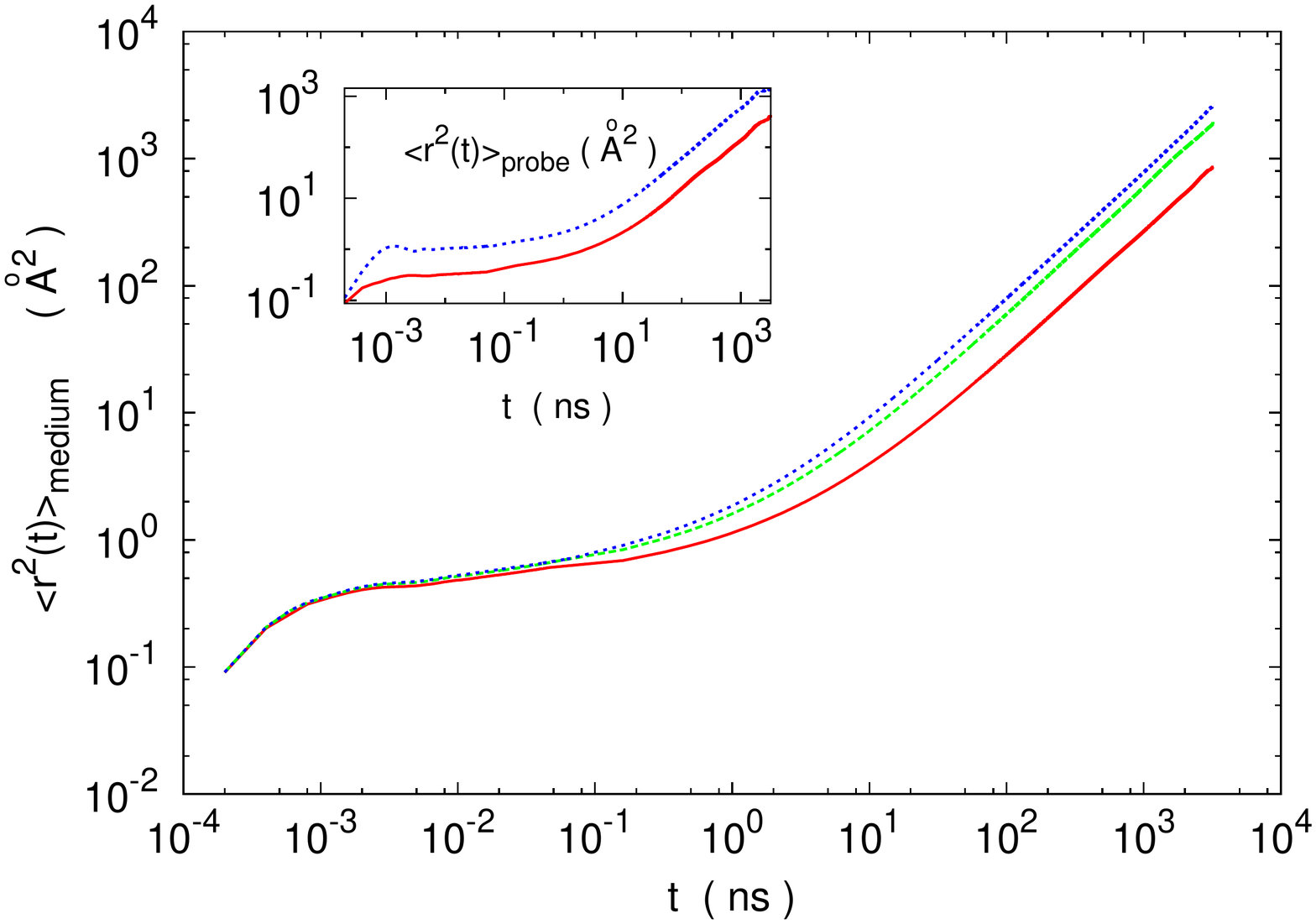}

{\em \footnotesize  FIG.3. (color online)  
Mean square displacement of (inset) the defects or (main Figure) the medium molecules, when the defects included are:
Small (blue dashed line), large (continuous red line), or we are in a pure liquid (dotted green line).T=120K.\\
}

Because the cages around the large probes are  smaller  than the medium's cages, and these probes are larger than the medium's molecules, we expect a decrease of the free volume around large probes.
 The decrease of the free volume around large probes then results in a decrease of the diffusion around these probes.
Similarly we expect an increase in the free volume around small probes resulting in an acceleration of the diffusive motions.

To verify that picture we evaluate the free volume $v_{f}$ around the probe from the height of the plateau of the mean square displacement of the probe (see the inset of Figure 3).
If $h$ is the plateau's height, we have $v_{f}\approx 0.75 h^{3/2}$.
Figure 3 shows that $h^{small} > h^{medium}>h^{large}$ ($h^{small}\approx1.06$\AA$^{2}$, $h^{medium}\approx 0.54$\AA$^{2}$, $h^{large}\approx 0.33$\AA$^{2}$) leading to the same relation for the free volumes, with  $v_{f}^{small} = 0.82$\AA$^{3} ; v_{f}^{medium} = 0.3$\AA$^{3}$ and $ v_{f}^{large} = 0.14$\AA$^{3}$.
The free volume for small probes is thus larger than for the mean medium molecules, leading to the increase of the diffusion for small probes that we observe in the inset of Figure 3.
As a result the inclusion of small or large molecules respectively increase or decrease the mean square displacements around the perturbation.
The defects thus slow down or accelerate the dynamics around them for large or small inclusions respectively.
The effect of these modifications of the local dynamics around the probes results in a modification of the whole dynamics as shown in Figure 4.
We see on the figure that the global diffusion coefficient follows the same trend and decreases for large probes (or increase for small ones).

We note that the modifications of the MSDs appear at the end of the plateau regime, around the characteristic time $t^{*}$ for which the cooperative motions are maximum, however the probes behave differently as shown in the inset of Figure 3.
For the probes the MSDs change at the beginning of the plateau.

We also note that this effect is temperature dependent and disappears at high temperature as shown in Figure 4.
The fragility, seen here as the diffusion coefficient variation from the pure exponential Arrhenius law $D=D_{0}e^{-E_{a}/k_{B}T}$, increases when large probes are included and decreases for small probes. As the increase of the activation energy when the temperature drops is usually explained by the increase of cooperative motions, that result suggests that the cooperativity increases for large probes and decreases for small ones. This result suggests that the cooperative motions increase when large probes are used and decreases for small probes.
We will see below that the DHs  follow that behavior.

\includegraphics[scale=0.33]{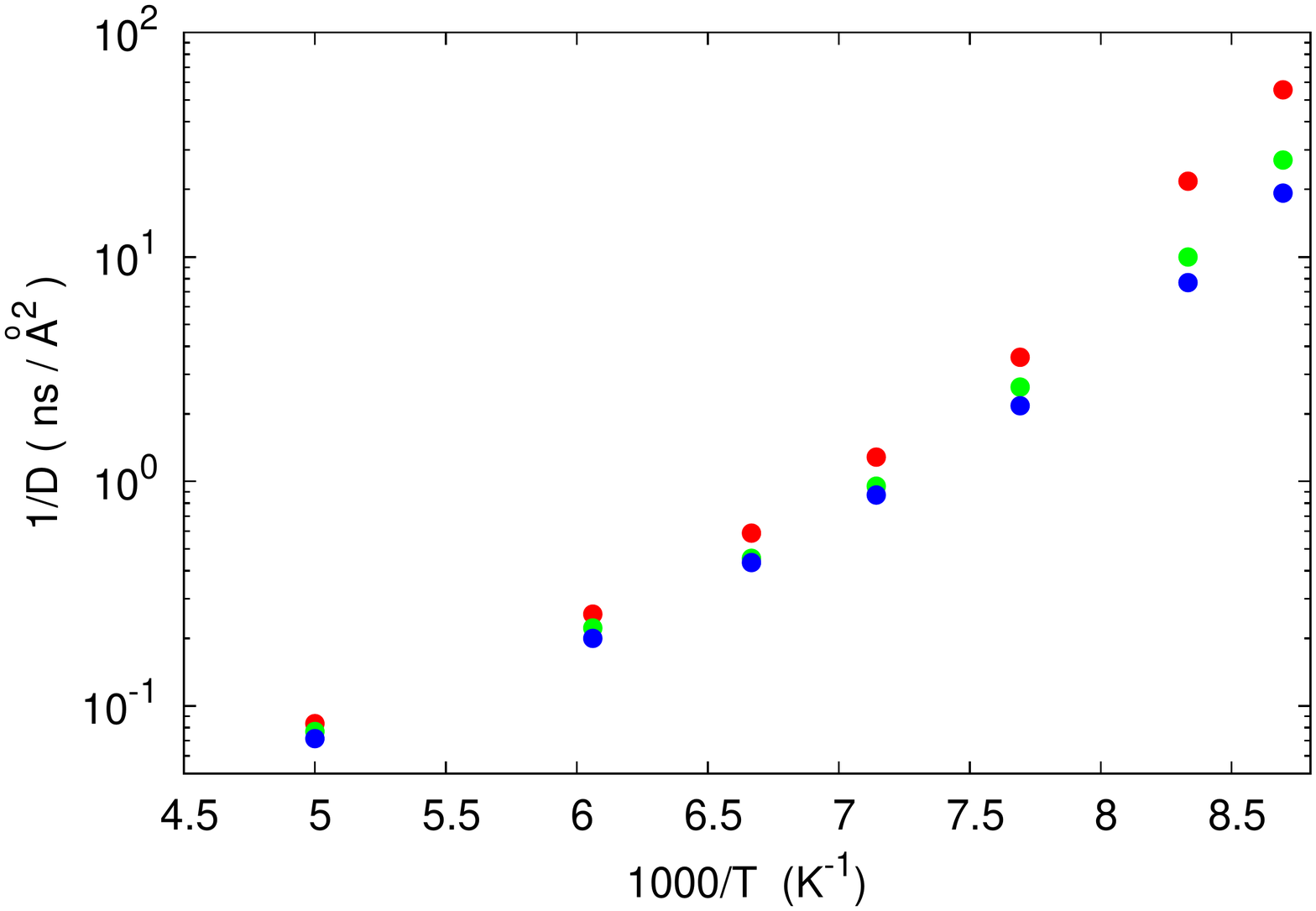}

{\em \footnotesize  FIG.4. (color online) Diffusion coefficient ($D$) for the medium molecules, versus temperature when small or large probes are included, compared with the diffusion coefficient of the pure liquid in the same conditions. From top to bottom: Large probes (red solid circles), pure medium (green solid circles), and small probes (blue solid circles). The medium molecules diffuse less when the probes are large.\\}


\includegraphics[scale=0.33]{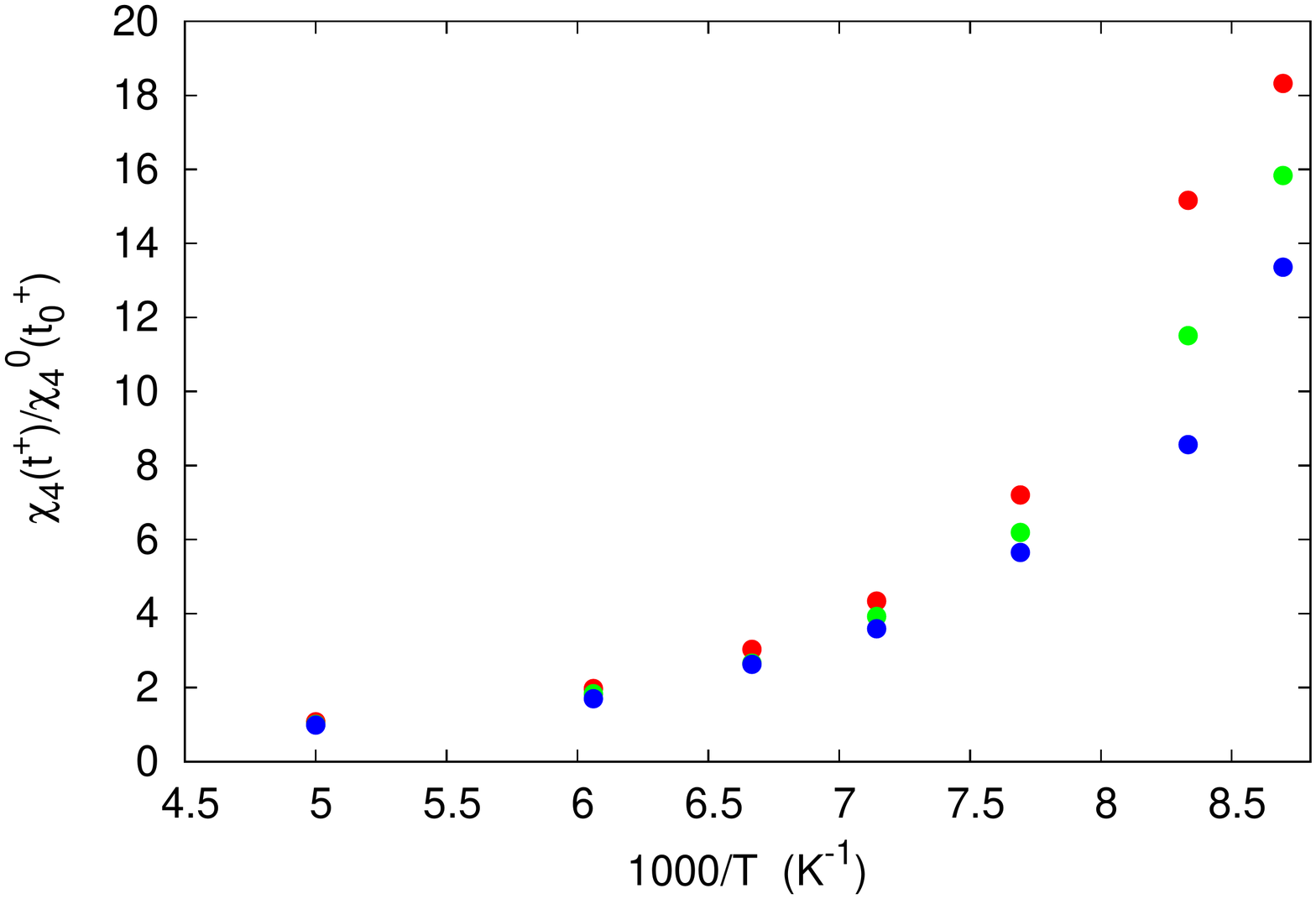}

{\em \footnotesize  FIG.5. (color online) Maximum value of the dynamic susceptibility $\chi_{4}(t^{+})$ for the medium molecules, versus temperature when small or large probes are included, compared with values from the pure liquid in the same conditions.  The function $\chi_{4}(t)$ reaches its maximum for $t=t^{+}$. From top to bottom: Large probes (red solid circles), pure medium (green solid circles), and small probes (blue solid circles). The dynamic susceptibility is normalized with its maximum value for the pure liquid at T=200K ($\chi_{4}^{0}(t_{0}^{+})$).\\}

\includegraphics[scale=0.33]{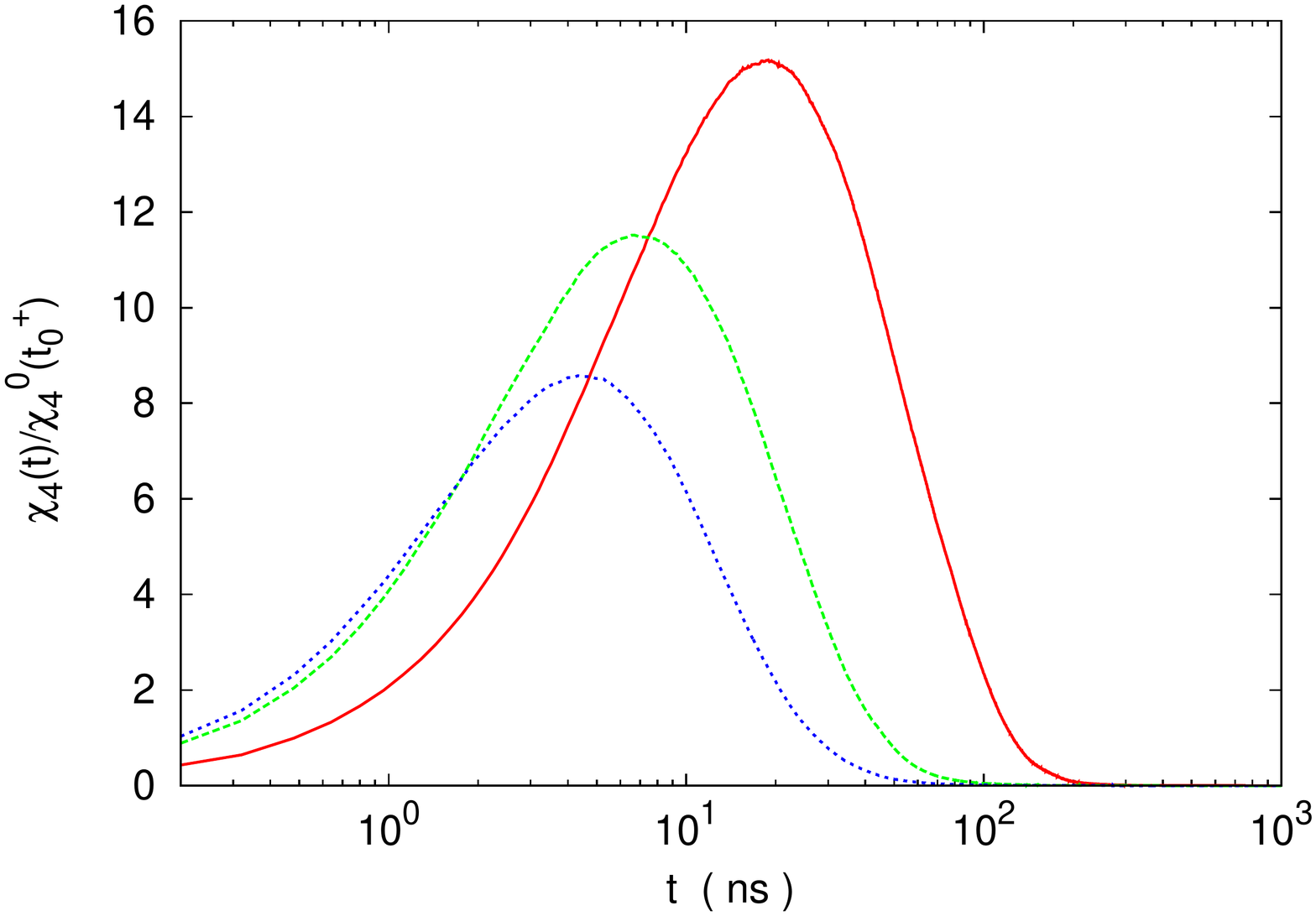}

{\em \footnotesize  FIG.6. (color online)  Dynamic susceptibility $\chi_{4}(t)$ for the medium molecules, versus time when small or large probes are included, compared with the dynamic susceptibility of the pure liquid in the same conditions. From top to bottom: Large probes (red continuous line), pure medium (green dotted line), and small probes (blue dashed line). T=120 K.
The dynamic susceptibility is normalized with its maximum value for the pure liquid at T=200K ($\chi_{4}^{0}(t_{0}^{+})$).\\}

The 4-point dynamic susceptibility\cite{dh1} measures the autocorrelation of the fluctuations of the mobility and as such is an efficient measure of the dynamic heterogeneity. We calculate the  dynamic susceptibility $\chi_{4}$ from the equation\cite{dh1}:\\

\begin{equation}
\chi _{4}(a,t)=\frac{\beta V}{N^{2}}\left( \left\langle
C_{a}(t)^{2}\right\rangle -\left\langle C_{a}(t)\right\rangle ^{2}\right) 
\label{e2}
\end{equation}
with 
\begin{equation}
C_{a}(t)={\sum_{i=1}^{N}{w_{a}}}\left( \left\vert {{{\mathbf{r}}}}_{i}(t)-{{{\mathbf{r}}}}_{i}(0)\right\vert \right) .  \label{e3}
\end{equation}

In these equations, $V$ denotes the volume of the simulation box, $N$ denotes the number of molecules in the box, and $\beta =(k_{B}T)^{-1}$. Also, the symbol $w_{a}$ stands for a discrete mobility window function, $w_{a}(r)$, taking the values $w_{a}(r)=1$ for $r<a$ and zero otherwise. We use the value $a=1.5$\AA $ $ in this work, which maximizes $\chi_{4}$ in our liquid at the density of the study. Note that some caution must be taken if small values of $a$ are chosen in the calculation (that may probe vibrational motions inside the cages instead of cage escaping motions) while larger values of the parameter $a$ lead to qualitatively similar results than with the parameter we chose.

\vskip0.5cm
Figures 5 and 6 show that the dynamic susceptibility is larger when we add large probes inside the medium, and smaller when we add small probes.
As the dynamic susceptibility measures the DHs, these results show that the DHs increase or decrease depending on the defects we add to the medium.
These results strongly suggest that the defects associated with the small molecules destroy the DHs, while the large molecules defects facilitate the DHs.
When we include defects (probes) inside the medium, 
Figure 5 shows that for both sort of probes the  DHs modification from the pure medium values, increases when the temperature drops.
Adding large probes to the medium, results in an increase of the DHs while  the dynamics slow down, following the same trend than the inverse of the diffusion coefficient in Figure 4.

\includegraphics[scale=0.33]{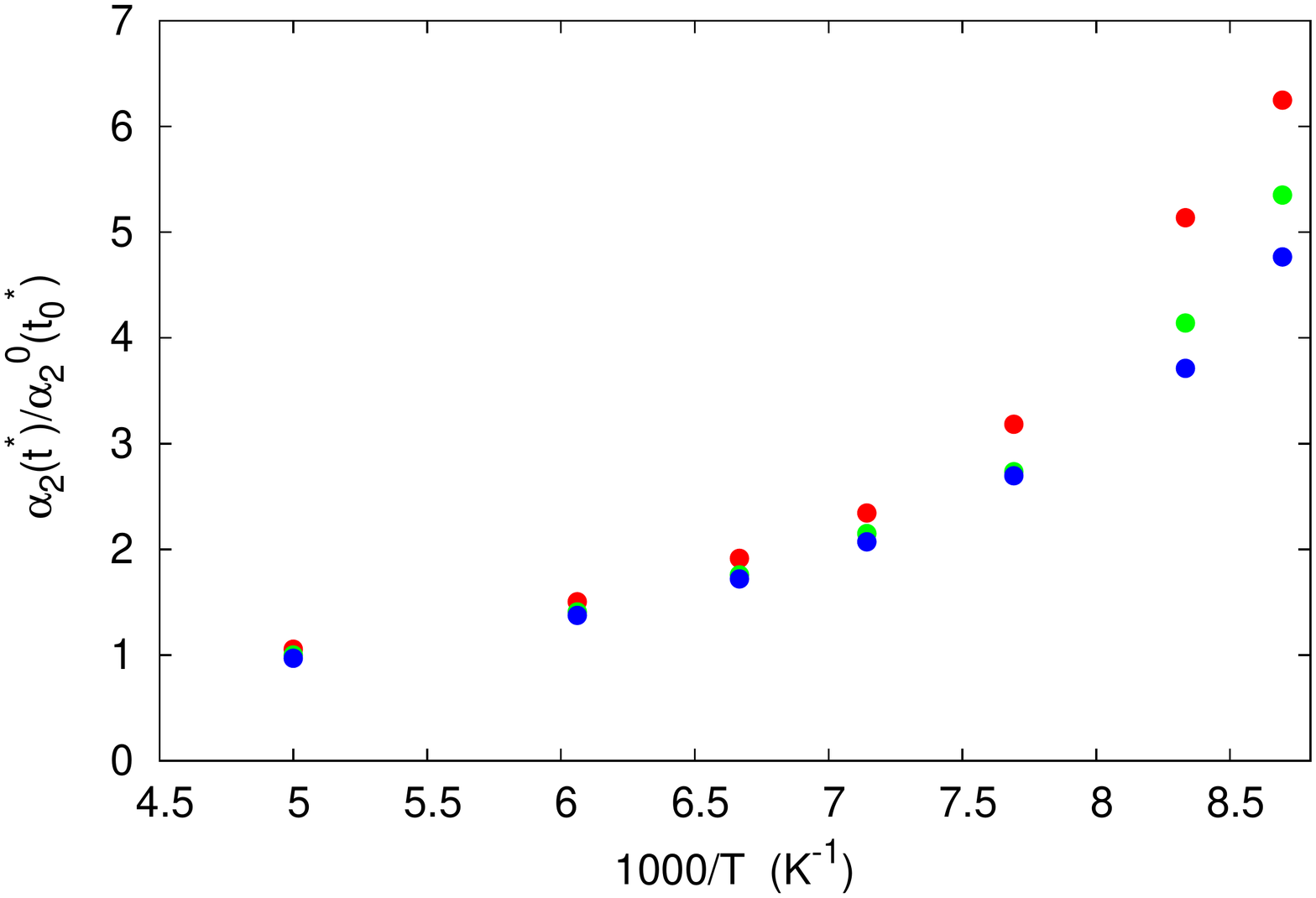}

{\em \footnotesize  FIG.7. (color online) Maximum value of the Non Gaussian parameter $\alpha_{2}(t^{*})$ for the medium molecules, versus temperature when small or large probes are included, compared with values from the pure liquid in the same conditions. From top to bottom: Large probes (red solid circles), pure medium (green solid circles), and small probes (blue solid circles).
The Non Gaussian parameter is normalized with its maximum value for the pure liquid at T=200K ($\alpha_{2}^{0}(t_{0}^{*})$).\\}

\includegraphics[scale=0.33]{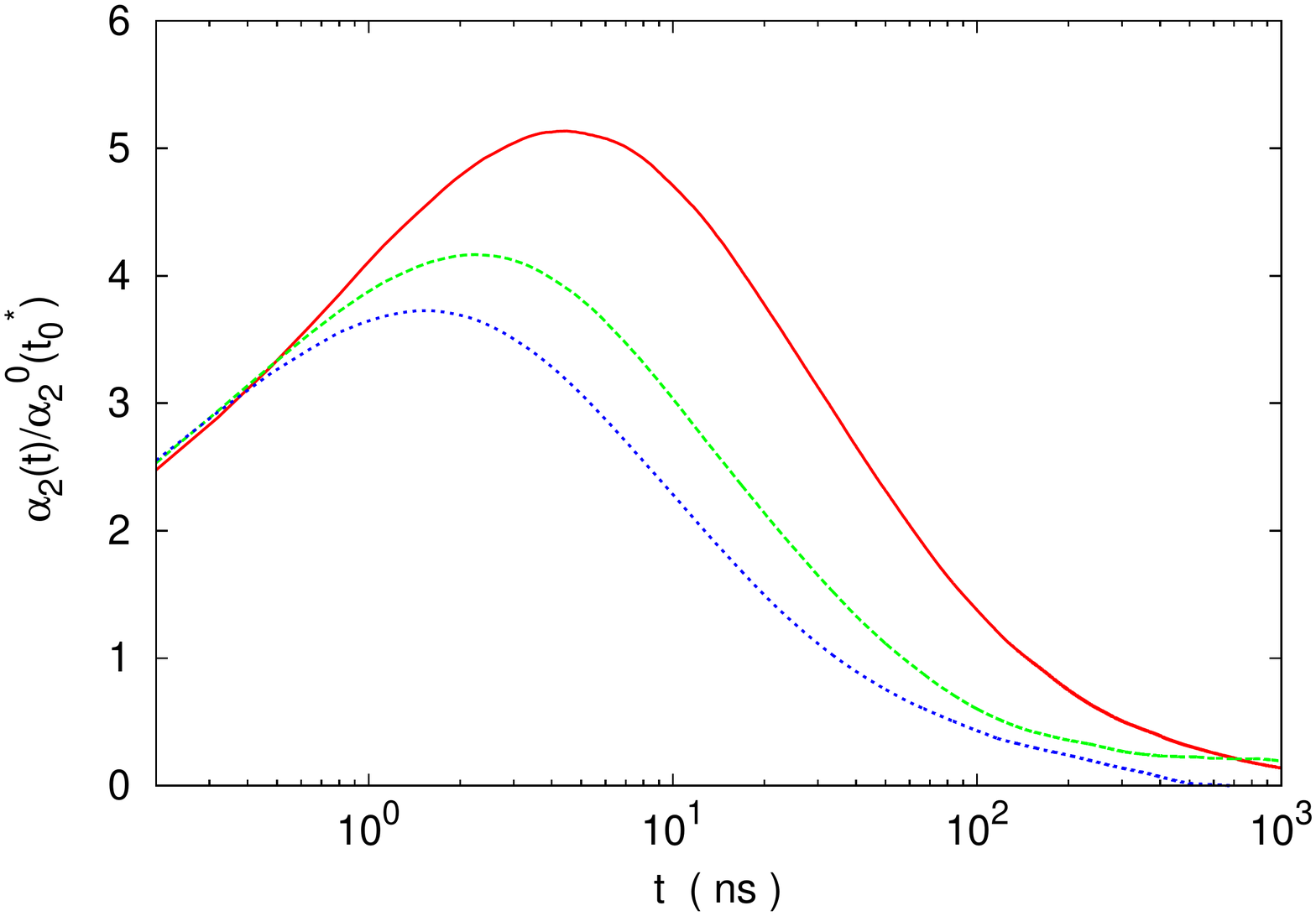}

{\em \footnotesize  FIG.8. (color online)  Non Gaussian parameter $\alpha_{2}(t)$ for the medium molecules, versus time when small or large probes are included, compared with the Non Gaussian parameter of the pure liquid in the same conditions. From top to bottom: Large probes (red continuous line), pure medium (green dotted line), and small probes (blue dashed line). T=120 K.
The Non Gaussian parameter is normalized with its maximum value for the pure liquid at T=200K ($\alpha_{2}^{0}(t_{0}^{*})$).
\\}

We display in Figure 7  the Non Gaussian parameter (NGP) $\alpha_{2}$ evolution with temperature, and in Figure 8 its time variation at a temperature T=120K.

The Non-gaussian parameter $\alpha_{2}(t)$ is defined as:

\begin{equation}
\alpha_{2}(t)=\frac{3 <r^{4}(t)>} {5<r^{2}(t)>^{2}}    -1  \label{e4}
\end{equation}

The NGP $\alpha_{2}(t)$ measures the  variations from the Gaussian shape of the self Van Hove correlation function,  that is predicted by Brownian motion.
As one of the signature of the dynamic heterogeneity is the appearance of a tail in the Van Hove originating from fast cooperative motions,  $\alpha_{2}(t)$ measures the dynamic heterogeneity.
We see in Figures 7 and 8 an evolution of $\alpha_{2}$ with temperature and with the probe  that is  qualitatively similar than the evolution we have observed for the dynamic susceptibility. At high temperature the probes have no effect on the DH and the circles merge in Figures 7 and 5 for $T\approx 200 K$. Then, as the temperature decreases the different probes lead to increasingly different DHs. The small probes decrease the DH while the large probes increase them.
We observe a smaller variation of  the Non Gaussian parameter than of the dynamic susceptibility when decreasing the temperature in our liquid.
Note that while the characteristic time for the susceptibility depends on the choice of the parameter $a$, the characteristic time of the NGP $t^{*}$ is an important characteristic of the supercooled liquid at the temperature of study. However we observe in both cases (Figures 6 and 8) that the maximum value of the NGP and of the susceptibility are shifted to larger times for large probes and to smaller times for small probes.

\includegraphics[scale=0.33]{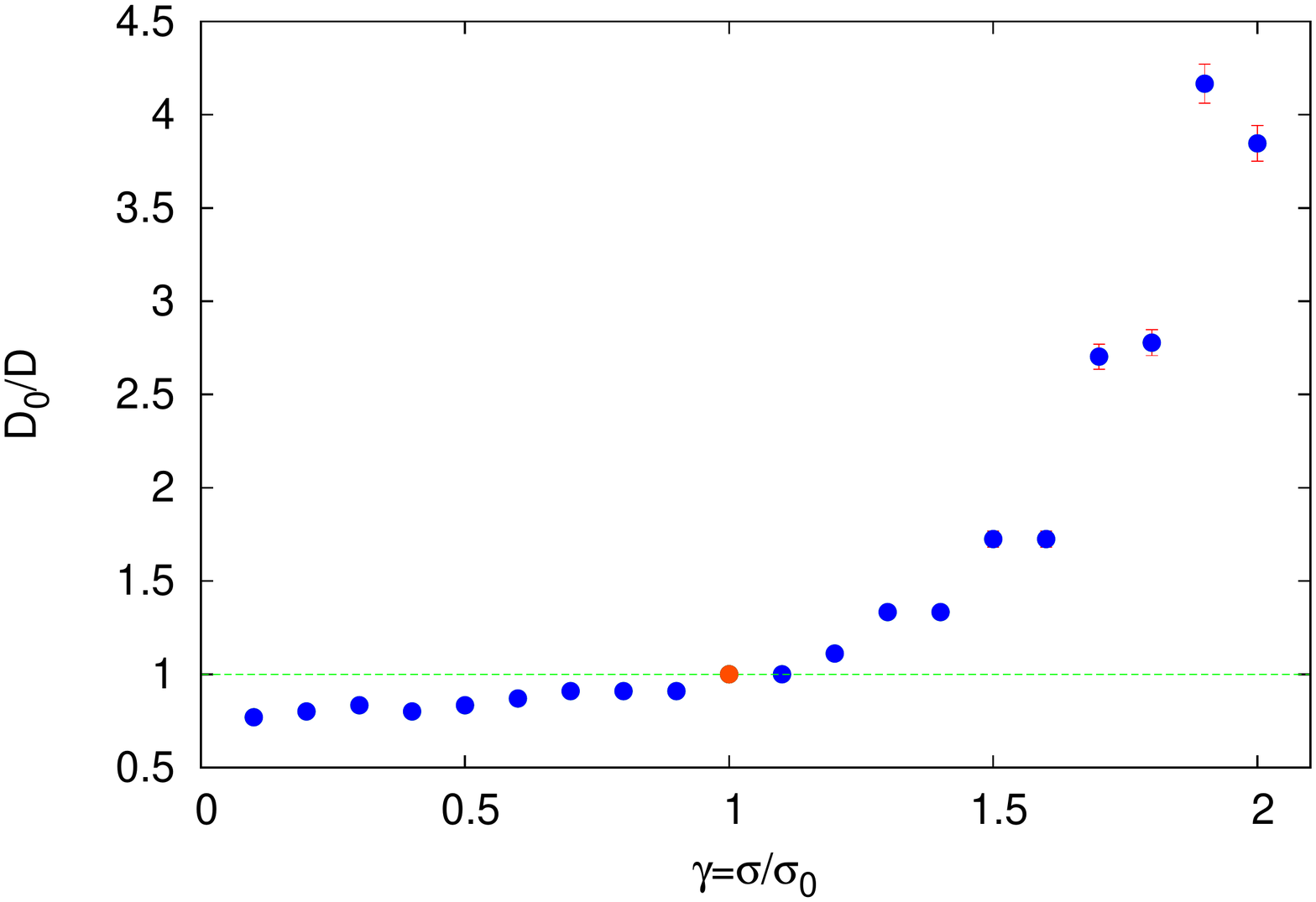}

{\em \footnotesize  FIG.9. (color online)  Inverse of the diffusion coefficient $1/D$ for the medium molecules normalized with the diffusion coefficient  $D_{0}$ of the pure liquid in the same conditions, versus the relative size $\gamma$ of the probe. T=120K. The red (dark) circle corresponds to the pur medium. This point was obtained with high accuracy.\\}

\includegraphics[scale=0.33]{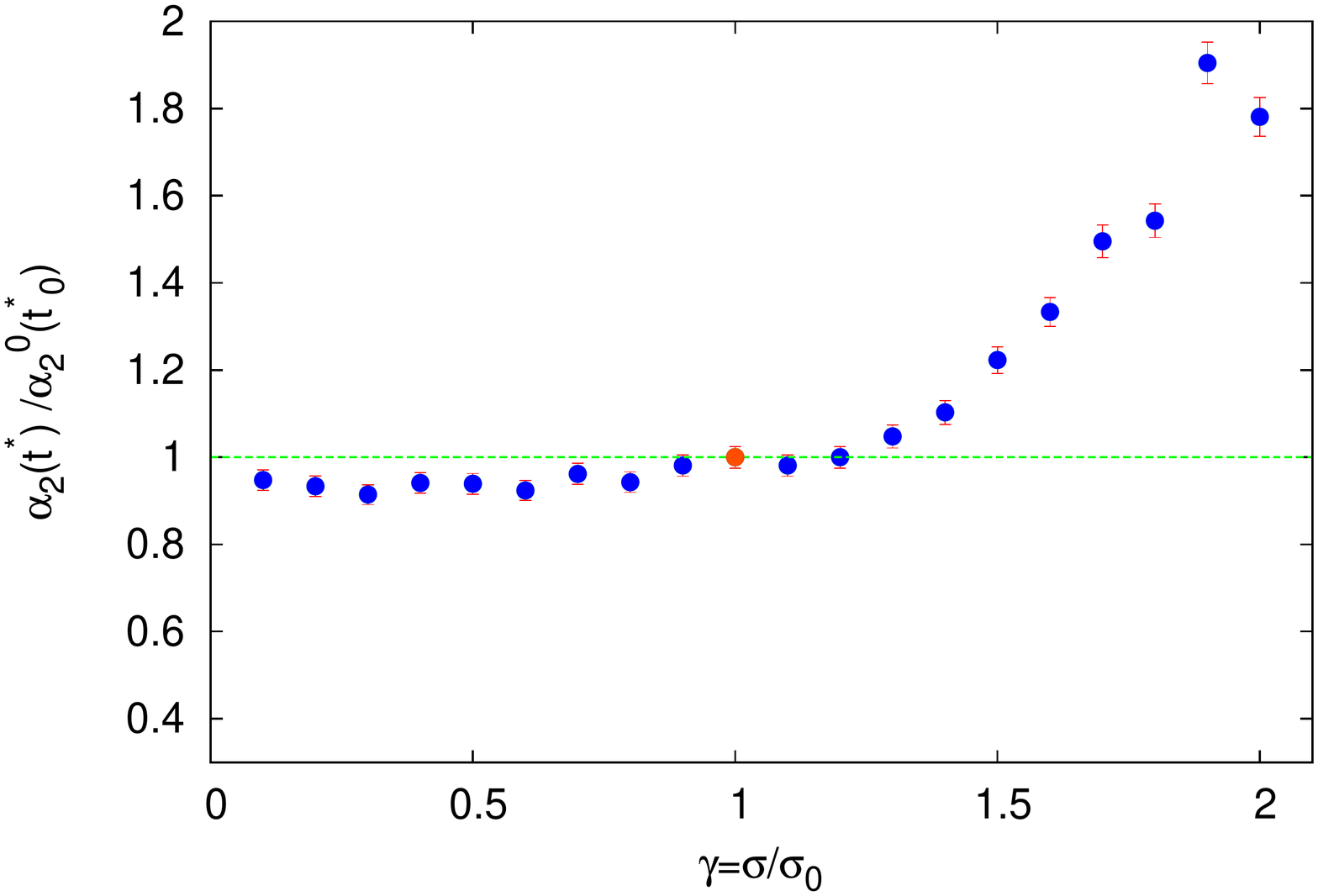}

{\em \footnotesize  FIG.10. (color online)  Maximum value of the Non Gaussian parameter $\alpha_{2}(t^{*})$ for the medium molecules, normalized with the Non Gaussian parameter  of the pure liquid $\alpha_{2}^{0}(t_{0}^{*})$ at the same temperature T=120K, versus the relative size $\gamma$ of the probe. The red (dark) circle corresponds to the pur medium. This point was obtained with high accuracy. Note that the temperature for the normalization (T=120 K) is here different from that of the normalization in Figures 7 and 8 (T=200K).\\}

We have studied the effect of two particular probes (a small probe and a large one) diluted inside a supercooled medium.
We will now show that the  effects observed are not qualitatively particular to the probes we chose previously.
For that purpose we will now vary the probe size and study the resulting effects on the dynamics and heterogeneity.
In the following, we use a probe that is similar to the medium molecule but with a size that has been multiplied by the factor $\gamma$ (i.e. $\sigma_{ij}^{probe}=\gamma  \sigma_{ij}^{m}$ and $d^{probe}=\gamma d^{m}$).
We show in Figures 9 and 10  the evolution of the diffusion coefficient $D$ and of the non gaussian parameter $\alpha_{2}$ with the relative size of the probe $\gamma$.
The diffusion coefficient decreases strongly when the size of the probe is larger than the medium molecule ($D_{0}/D$ increases in the Figure).
Larger probes result in smaller diffusion coefficients. Simultaneously we observe in Figure 10 an increase of the Non Gaussian parameter for large probes.
For probes smaller than the medium molecules, the diffusion is larger than for the pure medium, and roughly constant.
The Non Gaussian parameter is also roughly constant and smaller than the bulk value.

\vskip 1cm
\section{Conclusion}

In this work, using molecular dynamics simulations we have investigated the effect of small packing defects on the dynamic heterogeneity and diffusion processes in a supercooled liquid.
Our objective was to test the hypothesis of dynamic heterogeneities created by packing defects and the resulting modification of the liquid's dynamics. A secondary purpose was to better understand how diluted molecules modify the viscosity of a liquid below $T_{m}$.
 For these purposes we have used a simple diatomic glass-former with a few (1$\%$) probe molecules that were similar to the medium molecules but of a different size. Due to the simplicity of the molecules we were able to access very long time scales (of the order of the micro-second).
We found that the induced defects do not modify the dynamics at high temperature but below $T_{m}$ the modification increases rapidly when the temperature drops. This result shows that the effects are linked to the physics of the glass-transition.
We found that the relative size $\gamma$ between the diluted probe and the medium molecule strongly determines the dynamics when $\gamma >1$, but not when $\gamma <1$.
When $\gamma <1$ the cooperativity decreases and the dynamics is accelerated, while when $\gamma >1$ we observe the opposite effect.
In summary, our results show that a few (1$\%$)  packing defects  are able to strongly modify the dynamical heterogeneity
in a supercooled liquid, in most cases increasing the heterogeneity and in some cases destroying them.
These results come in support of the hypothesis of dynamic heterogeneities created by packing fluctuations in supercooled liquids.
Similarly, the packing defects also strongly affect the viscosity of the liquid, with larger defects leading to larger effects.

\end{document}